\documentclass[sigconf]{acmart}
\pdfoutput=1

\AtBeginDocument{%
  \providecommand\BibTeX{{%
    \normalfont B\kern-0.5em{\scshape i\kern-0.25em b}\kern-0.8em\TeX}}}

\usepackage{graphicx}
\usepackage{subcaption}
\usepackage{siunitx}
\usepackage{natbib}

\usepackage{tikz}
\usepackage{booktabs}

\usetikzlibrary{shapes,decorations,arrows,calc,arrows.meta,fit,positioning}
\tikzset{
    -Latex,auto,node distance =0.6 cm and 0.6 cm, thick, line width = 1.5,
    state/.style ={circle, draw, thick, minimum width = 0.8 cm, line width=1pt},
    point/.style = {circle, draw, inner sep=0.04cm,fill,node contents={}},
    bidirected/.style={Latex-Latex,dashed},
    el/.style = {inner sep=2pt, align=left, sloped},
    every picture/.style={line width=1pt}
}
\copyrightyear{2020}
\acmYear{2020}
\setcopyright{rightsretained}
\acmConference[ACM CHIL '20]{ACM Conference on Health, Inference, and Learning}{July 23--24, 2020}{Toronto, ON, Canada}
\acmBooktitle{ACM Conference on Health, Inference, and Learning (ACM CHIL '20)}

\begin{document}

\title{Machine Learning in Population and Public Health}

\author{Vishwali Mhasawade}
\email{vishwalim@nyu.edu}
\affiliation{New York University}
\author{Yuan Zhao}
\email{yuan.zhao@nyu.edu}
\affiliation{New York University
}
\author{Rumi Chunara}
\email{rumi.chunara@nyu.edu}
\affiliation{New York University}

\begin{abstract}
Research in population and public health focuses on the mechanisms between different cultural, social, and environmental factors and their effect on the health, of not just individuals, but communities as a whole. We present here a very brief introduction into research in these fields, as well as connections to existing machine learning work to help activate the machine learning community on such topics and highlight specific opportunities where machine learning, public and population health may synergize to better achieve health equity.
\end{abstract}

\maketitle
\section{Population and Public Health}\label{intro}
\begin{figure}
    \centering
    \includegraphics[scale=0.3]{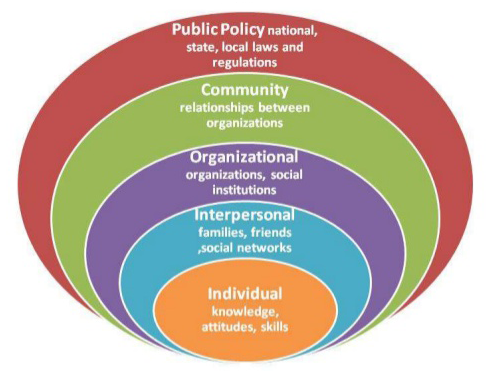}
    \caption{The socioecological model of health.}
    \label{fig:socio}
\end{figure}

Population and public health are approaches to health research and practice which aim to to understand what makes and keeps people healthy \cite{rose1985sick}. The major underpinning principle for this approach is that of \emph{health equity}, defined as \emph{``Minimizing avoidable disparities in health and its determinants -- including but not limited to health care -- between groups of people who have different levels of underlying social advantage or privilege, i.e., different levels of power, wealth, or prestige due to their positions in society relative to other groups''} \cite{braveman2006health}. Thus this guiding principle necessitates a focus on determinants, antecedents and other factors related to health outside the hospital. As further described by the socioecological model of health (Figure \ref{fig:socio}), a cornerstone of these domains which provides a conceptual model to illustrate how the health of an individual is affected by multiple factors operating at different levels in a hierarchy, these multi-level factors include public policies at the national and international level, availability of health resources within a neighborhood, community behavior, and ultimately the habits and behavior of individuals \cite{bronfenbrenner1977toward}. Understanding the complex interactions between individuals and their environments is crucial to realize not just how the health of the high-risk individuals, for example, those suffering from cardiovascular disease risk, can be improved but also what policies would benefit the community as a whole, such as will introducing healthier food options in a neighborhood help people to improve their diet \cite{stern2016people}? 

\subsection{Potential impact}
Realizing these complex interactions and identifying their effect on health outcomes remains one of the most prominent challenges as impact of doing so would be very large. For example, the inequalities in life expectancy for women with different income levels in the United States are large and growing \cite{national2015growing}. As well in the United States, social factors account for 25-60\% of deaths in any given year according to results from various meta-analyses \cite{heiman2015beyond}, and these factors play a significant role in prevention of the five leading causes of death (diseases of the heart, cancer, unintentional injuries, cerebrovascular diseases, and chronic lower respiratory diseases) \cite{centers2014up}. Worldwide, 80\% of the growing burden of noncommunicable diseases could be prevented through modifying behaviors such as reducing tobacco/alcohol or fat and salt consumption, promoting physical activity and improving environmental conditions such as air quality and urban planning \cite{world20092008}.\\

\par
Although in the past decade statistical and machine learning approaches have made considerable progress in automating clinical tasks, here we aim to illustrates where and how such techniques can be useful in a more holistic view of health. Towards this, we summarize existing challenges in public and population health and related opportunities. The outline of the document is as follows: we introduce the data used in population and public health studies in Section \ref{data}, followed by approaches for analyzing risk factors and disparities in Section \ref{equity}, and finally, present some directions for future work that leverages synergies in machine learning and population and public health in Section \ref{conclusion}.

\section{Data in public health}\label{data}
Considering that health is affected by individual-level and community-level attributes, a large focus of work is on identifying and understanding all of these potential factors and how they affect health outcomes. Naturally, this involves assessing not just biological factors but also social factors such as income, education, and socio-economic status (SES). The World Health Organization (WHO) defines these social determinants of health (SDoH) as \emph{``conditions in which people are born, grow, live, work and age'', such circumstances are shaped by the distribution of power, money, and resources, not just at the local level but also global and national levels} \cite{who2008closing}. It is important to note that these factors can come into play and have effect throughout the life course or across generations \cite{cable2014life}. For example, adverse childhood experience of parents was found to be associated with higher odds of poor health of the child \cite{le2018intergenerational}. Thus it is paramount to look beyond the individual's immediate risk factors and assess such inter-generational effects as well as the social, economic, and cultural environments that an individual identifies with. We next present some of the approaches to identifying and assessing social factors. Following we describe, through the deep literature on social determinants, common approaches for their measurement, and we distill challenges which can illuminate potential opportunities for machine learning.

\subsection{Measuring social determinants of health}
Traditional approaches to data collection in public health involve aggregation across multiple levels. Individuals often report data to public health practitioners and healthcare workers, which is then aggregated by local officials and forwarded to health ministries at the national level and finally analyzed at the international level by organizations like World Health Organization, United Nations. This results in robust, denominator-based data available from public health or governmental organizations, such as the national health and nutrition examination survey (NHANES) \cite{national1994plan} or the behavioral risk factor surveillance system (BRFSS) \cite{stein1993behavioral}.  These types of data systems also aim to capture different indicators which compose each construct (e.g. housing quality can be measured through data on rental status, sanitation status, crowding, indoor air quality, etc.) \cite{kusnoor2018collection}. However, there is also a loss of critical information at an individual level due to aggregation as well as temporal delays in the process. Accordingly, the rising ubiquity in technologies such as smartphones and social media sites like Twitter or Instagram has come to the forefront, making it possible to access high-resolution data that does not suffer from recall or information biases, and to capture information across daily behavioral patterns \cite{mhasawade2020role, chunara2017denominator, zhan2018using}. \\

\noindent These new data sources provide opportunity for measuring social determinants, but also come with their own challenges. After over a decade of their use in health studies, it is now possible to distill common challenges. Beyond data privacy, which we highlight but do not discuss here at length as there are several other reports which focus on it \cite{o2011informational, mooney2018big}, one of the most prominent issues is understanding the study's denominator, i.e., who all were included in study \cite{chunara2017denominator}. Moreover, since studies are often focused on specific groups in the community, this raises internal and external validity issues. First, it is important to understand what the measured variables are indicative of, and what constructs they represent. For example, recent work on how individual-level syndromic reports or passive data from individuals relates to microbiologic confirmation \cite{daughton2020comparison}, aims to address early challenges highlighted in using data such as Google searches to predict influenza \cite{lazer2014parable}. Research on the data generating process of new digital data sources is also imperative to understand what the data represents, and why data is being shared by individuals \cite{mhasawade2020population}. Second, can the results be extended to other situations, groups, or events; which sub-populations are representative of the hypothesis \cite{mitchell2004research}? To understand this phenomenon, a recent study analyzed the characteristics of different surveillance approaches (such as the questions asked, time of data collection) for influenza and their effect on predictive performance. Even if similar syndromic data is collected across all the approaches, there are considerable differences in the predictive value of the syndromic data \cite{chunaradiversitysur}, highlighting that the same type of data can represent different factors across studies (in this case the mode of data collection can affect the specificity/sensitivity of respiratory infection syndromic data). Although there are significant differences in sample representation and predictivity across studies, it is crucial to understand external validity as it provides a way to understand population-level characteristics that remain invariant across studies \cite{chunara2017denominator,mhasawade2020population}.\\

\noindent Challenges in measuring social determinants echo recent work in the computer science literature. Holstein et al. interviewed data scientists and synthesized findings around the importance of better data for improving the performance of machine learning models opposed to model development \cite{holstein2019improving}. Further, mapping from the construct to the observed space can itself be a place where bias can enter \cite{friedler2016possibility}, and thus special attention to which data and from whom it is gathered, should be highlighted in health research.

\subsection{Integrating social determinants of health in machine learning models}
\noindent 
Beyond work capturing social determinants using machine learning from person-generated sources \cite{akbari2019using, quisel2016intra}, a recent systematic review analyzed how social determinants have been used to study the risk factors of cardiovascular diseases \cite{yuan2020machine}. While common social determinants like age, gender, race, income, and education have been analyzed for estimating cardiovascular risk, most commonly the factors considered are at the individual-level even though research has clearly shown that community-level factors such as a person's neighborhood's overall income can also affect their disease risk. 
\cite{yuan2020machine}. Moreover, most studies to-date using machine learning models have involved associative studies. However, it is essential to understand the complex causal pathways between the social factors and the health outcomes in order to design effective interventions. Thus, there is a need to look beyond familiar data sources such as electronic health records (EHR), and develop approaches for evaluating the causal underpinnings in the data that captures these multi-level factors. Recognizing and understanding these social determinants of health across high dimensional multi-modal sources is an area which may benefit through the use of machine learning.

\subsection{Are social variables intervenable?}
Sometimes the use of social variables in causal models is restricted, under the premise that they are non-manipulable, or not intervenable \cite{glymour2014commentary}. Moreover, causal methods often assume stable unit treatment value (SUTVA). This implies that there is no interference and only one version of treatment, but this is often impossible with the complex interaction between social determinants. One proposed solution for this is to manipulate downstream mediators, such as encouraging children to read in order to increase their cognitive ability. But in order to make structural changes, we still need to address the root cause of disparities, and also aim at intervening on upstream social determinants, such as improving SES \cite{lehman2019addressing}.

At the same time, quantitative health researchers have aimed to understand how structural factors relate to health while also keeping in mind what is possible to intervene. One example of this is a simulation study aimed at understanding factors related to reducing the prevalence of chronic illnesses. While it was postulated that multiple social factors like social cohesion, jobs/income, education level, individual behavior, housing and healthcare interventions can lead to reduction in chronic illness, the simulation study narrowed down which factors were related based on data from their setting. They were also able to quantify the intervention needed for each social factor in order to lower chronic illness prevalence \cite{mahamoud2013modelling}.

\section{Types of health tasks}\label{equity}
Guided by the principle of health equity defined above, we see that a broad focus on multi-level aspects related to health in order to understand as well as intervene on is essential. Accordingly, we developed a taxonomy to consolidate types of tasks. We provide examples of where machine learning has been used in each type of task, and where further innovation in each of these is possible as well. The taxonomy includes: 1) identification of factors (biological, environmental, social, etc.) and their relation to health, 2) design/evaluation of interventions and policies on health, 3) prediction of outcomes, and 4) allocation of resources at the individual or group level. We briefly describe examples of such tasks below, full discussion on the health taxonomy can be found at \url{https://chunaralab.github.io/MLPH}. 
\begin{enumerate}
    \item \textit{Identification of factors}. Learning what contributes to health outcomes is a common theme across biological, social and other factors, and a good opportunity for learning from data which is a fundamental aspect of machine learning. The concept of identification comes into play across a broad range of studies in health, from learning biological mechanisms \cite{burgess2018inferring}, assessment of treatment effects \cite{lodi2019effect} to epidemiological studies of spatial factors \cite{bhatt2013global}.
    \item \textit{Design of interventions}. Interventions can occur at multiple levels and through different mediums, providing diversity in this category as well. For example, group based intervention programs are one of the means for reducing substance abuse by reinforcing positive behavior. With the advent of digital data available, social networks can be used for designing interventions and targeting high risk individuals in proximity with already exposed users. For example, \citet{rahmattalabi2018influence} present an influence-based partitioning of social networks to identify high impact intervention groups.
    \item \textit{Prediction of outcomes}. While predicting treatment effectiveness \cite{kreif2015evaluation}, mortality risk \cite{rajkomar2018scalable}, hospital readmission \cite{galiatsatos2019association}, and disease prognosis \cite{dugan2015machine} are some tasks well studied in machine learning, predicting risk scores with clinical algorithms while mitigating health disparities as well as prediction of outside-hospital events are still crucial challenges \cite{vyas2020hidden}.
    \item \textit{Allocation of resources}. Resource allocation is a well studied problem in artificial intelligence. For example, in order to ensure equity in access to resources \citet{snyder2018organ} suggest an allocation system combining a medical priority score and a geographic feasibility score to facilitate organ allocation across geographical boundaries. However, there is further opportunity to consider the types of resources and attributes which are considered in allocation, with a health equity lens. 
\end{enumerate}

\subsection{Causal approaches to understand mechanisms between social determinants and health}
Causal approaches are critical for understanding the mechanism between different social factors and health outcomes. Causal methods provide a way to incorporate prior knowledge about mechanisms into modeling, and also to identify what is intervenable. For example, it is known that the social construct of `race' has direct effects on the health of individuals belonging to the disadvantaged group. To improve the health of the marginalized individuals, it may be argued that \emph{race} cannot be intervened upon and there is nothing that can be done to \emph{race} to improve the existing scenario. However, the social construct can be decomposed into several interacting factors like parents' genetics, family culture, social perspectives about appearance, the early life socio-economic conditions, and late-life socio-economic conditions. While some of the factors like parents' genetics cannot be intervened upon, it is still possible to mitigate health disparities by focusing on intervenable factors like socio-economic conditions \cite{vanderweele2014causal}. \\

\noindent Another aspect of modeling social determinants apart from decomposing social constructs into interacting factors is realizing the different pathways and interaction effects (i.e. learning the data generating process). For example, an analysis of the health behavior of individuals on social media website Instagram revealed that the immediate environment comprising of the food resources in the neighborhood as well as the social network of an individual (the profiles being followed and interacted with) affect what individuals post on Instagram about dining \cite{mhasawade2020role}. What is interesting here is the focus on learning the mechanism through which the social environment has an effect on the health behavior, more broadly than existing public health knowledge. While existing work in public health has shown there is a direct effect of the availability of resources in the proximity of an individual and their health \cite{morland2009obesity}, this work augments current understanding to include factors from the online environment and understanding the mechanism between the online and social environment and in turn its effect on what an individual posts online.\\

\begin{figure}
    \centering
    \includegraphics[scale=0.3]{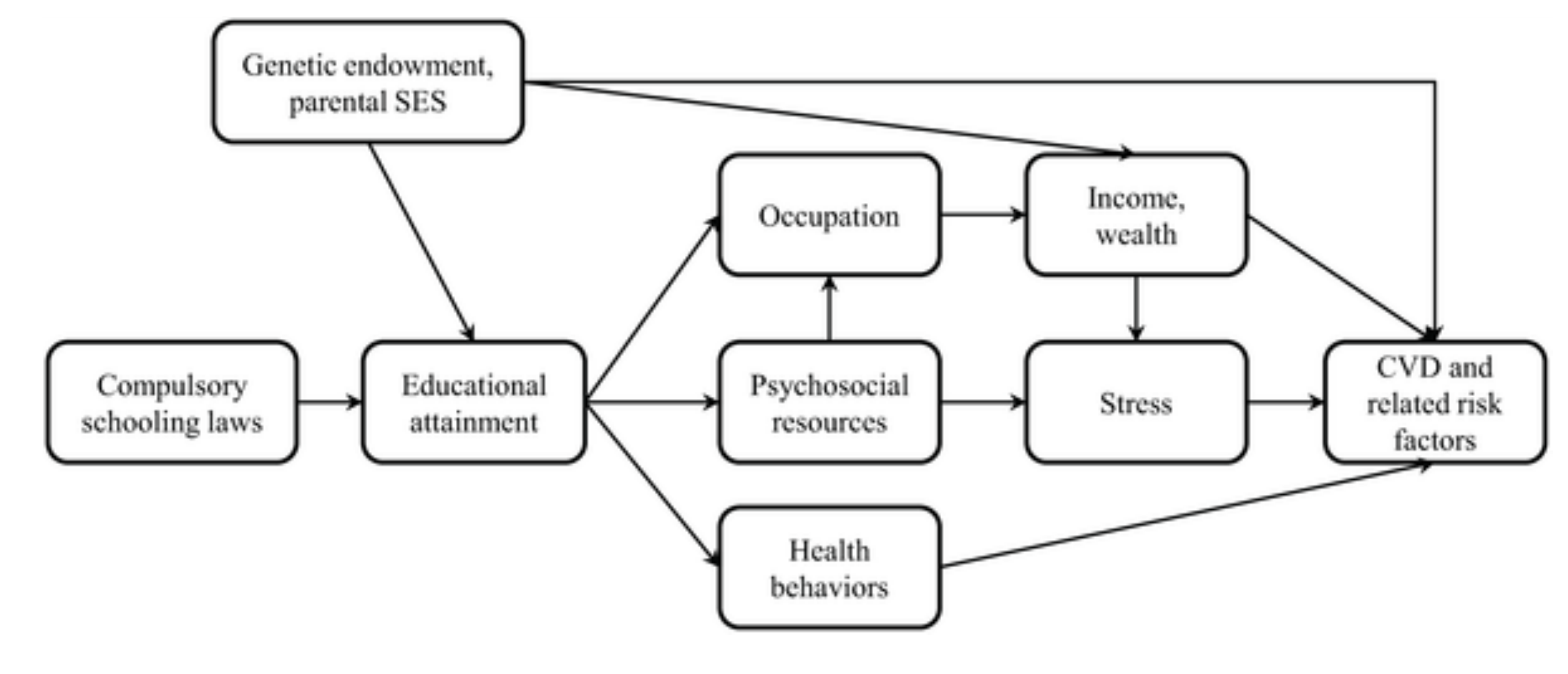}
    \caption{Conceptual model linking educational attainment level with cardiovascular disease (CVD) \cite{hamad2019educational}.}
    \label{fig:education_cvd}
\end{figure}

\noindent Causal methods are critical in models, including when considering social variables. A natural experiment across various states in the United States (with varying compulsory education laws) analyzed the impact of education level on cardiovascular disease risk. A postulated causal graph with multiple pathways from education level to cardiovascular disease risk is represented in Figure \ref{fig:education_cvd}. A simple association method concludes that education level decreases cardiovascular disease risk across all risk factors such as BMI, cholesterol, smoking, and depression. On the other hand, a causal approach, known as instrumental variables (IV), presents that education does not improve BMI and cholesterol risks \cite{hamad2019educational}. This can be attributed to the sedentary lifestyle resulting from improved income sources correlated with increased education levels. It is thus essential to identify the various interacting factors comprising a social construct and estimate the effects of social determinants via appropriate causal methods.

\subsection{Pitfalls with `proxies' in health modeling}
Although it is necessary to include and analyze the several comprising factors of social constructs like `race,' it is often unknown what the comprising factors are, how they interact, and how to model them. An inaccurate understanding of variations within and between groups, along with difficulty in attaining the relevant social factors have lead to the use of \emph{known} social constructs like race as \emph{proxies} for these unknown, unmeasured factors. A recent study highlighted several clinical tools across cardiology, nephrology, obstetrics, and many other specialties, which all use race while estimating risk factors, and how this use severely compromises the health of marginalized individuals \cite{vyas2020hidden}. A specific example of this is in the common way that kidney function is estimated using glomerular filtration rate (GFR) (estimated glomerular filtration rate) which uses serum creatinine levels and the race of the individual. Serum creatinine is the waste product in the blood resulting from muscle activity, is absorbed from the blood by the kidneys. However, under abnormal kidney function, the level of serum creatinine increases in the blood. Another reason for the increased serum creatinine level is higher muscle mass, which is attributed as the reasoning for considering the race factor. However, further research and understanding of the construction of this equation have illuminated that the use of race is not appropriate. It is not an accurate measure of the proxy, and furthermore results in delayed and disparate treatments to patients identifying as \emph{Black} individuals \cite{eneanya2019reconsidering,levey2020kidney,grubbs2020precision}. Discussion of these issues has led to the elimination of race considerations in calculating eGFR in several hospitals around the United States \cite{raceAKI,raceAKIdrop}. How any relevant genomic variation can be assessed and reported without stratifying populations based on factors like race and ethnicity is a challenge to be addressed \cite{bonham2018examining}. 

\subsection{Multiple axes of disparities and intersectionality}
\begin{figure}
    \centering
    \scalebox{0.8}{\begin{tikzpicture}
    \node[state,rectangle] (d) at (0,0) {Identify health determinants};
    \node[state,rectangle] (m) [below =of d] {Determine health outcomes};
    \node[state,rectangle] (a) [below =of m] {Measure health disparities};
    \node[state,rectangle] (i) [below =of a] {Design appropriate interventions};

    \path (d) edge (m);
    \path (m) edge (a);
    \path (a) edge (i);
    
\end{tikzpicture}}
    \caption{Pipeline for detecting health disparities}
    \label{fig:pipeline}
\end{figure}
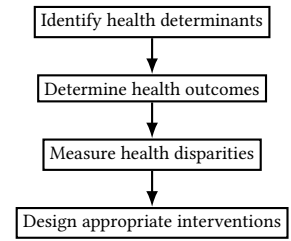

A health disparity/inequality is a particular difference in type of health where disadvantaged social groups like women, poor, racial/ ethnic minorities experience greater health risks \cite{braveman2006health}. Such disparities are reflective of social oppression and its influence on the health of disadvantaged individuals. It is essential to identify what leads to disparate health outcomes in order to design interventions to mitigate disparities and improve the health of high risk populations. This involves multiple tasks involving what factors affect health including both individual-level and social factors as mentioned before \cite{mhasawade2020role,mccradden2020ethical}, measuring the health outcomes \cite{chen2020robustly,obermeyer2019dissecting}, measuring disparities across social groups \cite{braveman2004approach,penman2016measurement,rajkomar2018ensuring}, and finally designing policies to mitigate the disparities \cite{rajkomar2018ensuring}. The steps towards mitigating disparities are summarized in Figure \ref{fig:pipeline}.\\

\noindent We now present a specific challenge in measuring disparities across groups of people. Health disparities are often measured by considering the average health outcomes of individuals identifying to a specific social construct say \emph{race}. For example, is the psychiatric readmission rate the same across racial groups \cite{chen2020robustly}. However, measuring health disparities as averages is not enough to identify the narrower sub-populations still suffering from the burden. A longitudinal study of prenatal care access for childbearing women in California across 1994-1995 and 1999-2001 reveals that the federal and state policies within these periods, on average, improved the health of pregnant women. However, disparities continued to exist across income groups even with the introduction of policies \cite{braveman2006health}. Thus, it is vital to measure differences across multiple social axes such as race, income, and education and primarily focus on individuals existing at the intersections of social disadvantages \cite{crenshaw1989demarginalizing}.\\

\noindent  One approach for addressing the disparities at intersections of social factors is a multilevel approach known as MAHIDA (Multilevel Analysis of Individual Heterogeneity and Discrimination Accuracy) \cite{evans2018multilevel}. It involves decomposing the total variance into a) between-strata variance focusing on identifying and assessing disadvantaged social groups, and b) within-strata variance which aims to identify individuals within a social group that are at added disadvantaged as compared to other members of the group. The approach presents several advantages over fixed-effect models that include interaction terms for multiple sensitive attributes. These include restricting the parameter growth to linear compared to geometric and adjusting for the sample sizes of the intersectionalities. The approach is consistent with the field of eco-epidemiology that cautions against aggregation. Studies have also focused on incorporating information across multiple environments to learn population-level characteristics, especially when subgroups are underrepresented across individual studies \cite{mhasawade2020population}.

\subsection{Health disparities and algorithmic fairness}
The growing use of machine learning methods in healthcare has raised the question of whether the model outcomes are discriminatory based on variables like race and ethnicity that are often used in constructing predictions \cite{chen2020robustly,mhasawade2020population}. A recent literature in machine learning and statistics aims on ensuring that model outcomes are not discriminatory towards individuals who have the same \emph{merit} \cite{kasy2020fairness,mitchell2018prediction}. For example, multiple fairness definitions like equality of opportunity, demographic parity have been suggested as approaches to restrict model outcomes to ensure fairness. However, given the complex causal relationships between the biological, social and environmental factors that lead to disparities in health outcomes \cite{mccradden2020ethical} outlined here, questions remain regarding such advances in algorithmic fairness and how they may interface with health disparities \cite{mccradden2020ethical, kasy2020fairness,friedler2016possibility}. Moreover, assessing disparate health outcomes becomes challenging when the underlying causal mechanisms are not known. We briefly outline specific challenges regarding algorithmic fairness and health disparities. \\

\noindent \textbf{Algorithmic design goals.} First, it is crucial to ensure that the algorithm design supports the goal of \emph{health equity}. For example, \citet{obermeyer2019dissecting} present that racial disparities in risk scores are a result of considering financial cost expenditure as a proxy for health care needs. It is therefore essential to be aware of such proxies and also assess disparities through different causal pathways, namely 1) direct, where the social construct has a direct impact on the outcome, and 2) through indirect causal pathways which provide an opportunity for performing interventions \cite{wu2019pc}.

\begin{figure}
    \centering
    \scalebox{0.9}{\begin{tikzpicture}
    \node[state] (e) at (0,0) {E};
    \node[state] (i) [below =of e] {I};
    \node[state] (n) [below =of i] {N};
    \node[state] (p) [right =of i] {P};
    \node[state] (h) [below right =of p] {H};
    \node[state] (c) [above right =of h] {C};
    
    \path (e) edge (p);
    \path (i) edge (p);
    \path (n) edge (p);
    \path (p) edge (h);
    \path (c) edge (h);
    
\end{tikzpicture}}
    \caption{Perceived protected attribute $P$ is composed of several social factors such as education $E$, income level $I$, and neighborhood characteristics $N$. Health outcome $H$ is to be predicted using the clinical variables $C$, and protected attribute $P$ while ensuring that the model prediction is fair with respect to $P$.}
    \label{fig:un_variance}
\end{figure}
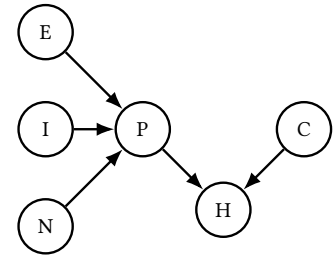

\noindent \textbf{Unexplained variance resulting from proxies}. Consider the scenario represented in Figure \ref{fig:un_variance} where we are concerned with predicting the health outcome, say risk for cardiovascular disease using the protected attribute, and clinical conditions. We also want to restrict models outcomes to be fair using some fairness metric like demographic parity. Even if we are successful in ensuring that the risk is fair with respect to the \emph{racial identity}, we are still left with unexplained variance for the social components of race. As we illustrate above, race is a social construct, and it would be unclear if such a model accounts for variances in the true mechanisms which race is acting as a proxy for. For example, do equal risk scores for Black and non-Black patients also ensure that there is no disparity across, say lower-income Black patients vs. higher-income Black patients. Thus, there is a need to both identify intersectional social groups as well as underlying causal mechanisms, and ensure algorithmic fairness for the same.

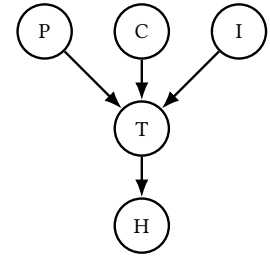
\begin{figure}
    \centering
    \scalebox{0.9}{\begin{tikzpicture}
    \node[state] (p) at (0,0) {P};
    \node[state] (c) [right =of p] {C};
    \node[state] (i) [right =of c] {I};
    \node[state] (t) [below =of c] {T};
    \node[state] (h) [below =of t] {H};

    \path (p) edge (t);
    \path (c) edge (t);
    \path (i) edge (t);
    \path (t) edge (h);
    
\end{tikzpicture}}
    \caption{Perceived protected attribute $P$, clinical variables $C$, and the insurance type $I$  used to determine the treatment $T$, and ultimately assess the health outcomes $H$.}
    \label{fig:insurance}
\end{figure}

\noindent \textbf{Equity vs. in-sample fairness}. \noindent Since data input for the algorithms may not be representative of the population for which decision might be made, it is crucial to be aware of the sub-populations included in the study \cite{yang2020causal,nishtala2020missed}, and mitigate disparate outcomes for under-represented sub-populations \cite{mhasawade2020population}. Figure \ref{fig:insurance} represents a common case in healthcare where different individual factors are used to make algorithmic decisions for patients. A fair algorithmic restricting the model decisions in favor of $P$ still suffers from population unfairness even if it achieves in-sample fairness. This is worsened by considering the insurance type of the individual, which leads to an association between the said insurance and health outcomes. It is important to note that often individuals are unable to access care, and remain at higher risk. Thus, it is imperative to consider if the aims of our efforts will alleviate and not exacerbate health inequities.

\section{Conclusion}\label{conclusion}
Health equity is a vast concept, and one that requires understanding and assessment in an ongoing manner, as factors related to health are shaped and change. Through the summary of existing work in population and public health, specific discussion of the importance of social determinants and challenges in their measurement and incorporation into causal models, we have synthesized the many open areas for \emph{machine learning}, to advance and build on research and practice in this area. The taxonomy helps lay out different areas in health where machine learning has and can play an important role; identifying factors related to health outcomes, design/evaluation of interventions and policies, prediction of outcomes and allocation of resources. 
Finally, we also discuss how the important growing research on algorithmic fairness interfaces with health disparities. Full discussion on the health taxonomy and related topics can be found at \url{https://chunaralab.github.io/MLPH}. In sum, there are many opportunities to build on the deep body of work on health equity. We hope that this work serves to inform and activate the machine learning community on these critical topics.

\section{Acknowledgments}
We thank Dr. Stephanie Cook and Harvineet Singh for helpful discussions on this topic. The work was partially supported under National Science Foundation grant 1845487.

\bibliographystyle{abbrvnat}
\bibliography{chil_tutorial.bib}

\begin{thebibliography}{64}
\providecommand{\natexlab}[1]{#1}
\providecommand{\url}[1]{\texttt{#1}}
\expandafter\ifx\csname urlstyle\endcsname\relax
  \providecommand{\doi}[1]{doi: #1}\else
  \providecommand{\doi}{doi: \begingroup \urlstyle{rm}\Url}\fi

\bibitem[Akbari and Chunara(2019)]{akbari2019using}
M.~Akbari and R.~Chunara.
\newblock Using contextual information to improve blood glucose prediction.
\newblock \emph{Machine Learning for Healthcare}, arXiv preprint
  arXiv:1909.01735, 2019.

\bibitem[Bhatt et~al.(2013)Bhatt, Gething, Brady, Messina, Farlow, Moyes,
  Drake, Brownstein, Hoen, Sankoh, et~al.]{bhatt2013global}
S.~Bhatt, P.~W. Gething, O.~J. Brady, J.~P. Messina, A.~W. Farlow, C.~L. Moyes,
  J.~M. Drake, J.~S. Brownstein, A.~G. Hoen, O.~Sankoh, et~al.
\newblock The global distribution and burden of dengue.
\newblock \emph{Nature}, 496\penalty0 (7446):\penalty0 504--507, 2013.

\bibitem[Bonham et~al.(2018)Bonham, Green, and
  P{\'e}rez-Stable]{bonham2018examining}
V.~L. Bonham, E.~D. Green, and E.~J. P{\'e}rez-Stable.
\newblock Examining how race, ethnicity, and ancestry data are used in
  biomedical research.
\newblock \emph{Jama}, 320\penalty0 (15):\penalty0 1533--1534, 2018.

\bibitem[Braveman(2006)]{braveman2006health}
P.~Braveman.
\newblock Health disparities and health equity: concepts and measurement.
\newblock \emph{Annu. Rev. Public Health}, 27:\penalty0 167--194, 2006.

\bibitem[Braveman et~al.(2004)Braveman, Egerter, Cubbin, and
  Marchi]{braveman2004approach}
P.~A. Braveman, S.~A. Egerter, C.~Cubbin, and K.~S. Marchi.
\newblock An approach to studying social disparities in health and health care.
\newblock \emph{American Journal of Public Health}, 94\penalty0 (12):\penalty0
  2139--2148, 2004.

\bibitem[Bronfenbrenner(1977)]{bronfenbrenner1977toward}
U.~Bronfenbrenner.
\newblock Toward an experimental ecology of human development.
\newblock \emph{American psychologist}, 32\penalty0 (7):\penalty0 513, 1977.

\bibitem[Burgess et~al.(2018)Burgess, Foley, and Zuber]{burgess2018inferring}
S.~Burgess, C.~N. Foley, and V.~Zuber.
\newblock Inferring causal relationships between risk factors and outcomes from
  genome-wide association study data.
\newblock \emph{Annual review of genomics and human genetics}, 19:\penalty0
  303--327, 2018.

\bibitem[Cable(2014)]{cable2014life}
N.~Cable.
\newblock Life course approach in social epidemiology: an overview, application
  and future implications.
\newblock \emph{Journal of epidemiology}, page JE20140045, 2014.

\bibitem[{CDC}(2014)]{centers2014up}
{CDC}.
\newblock Up to 40 percent of annual deaths from each of five leading us causes
  are preventable.
\newblock \emph{Atlanta, GA: Centers for Disease Control and Prevention}, 2014.

\bibitem[Chen et~al.(2020)Chen, Agrawal, Horng, and Sontag]{chen2020robustly}
I.~Y. Chen, M.~Agrawal, S.~Horng, and D.~Sontag.
\newblock Robustly extracting medical knowledge from ehrs: A case study of
  learning a health knowledge graph.
\newblock In \emph{Pac Symp Biocomput}, pages 19--30. World Scientific, 2020.

\bibitem[Chunara et~al.(2017)Chunara, Wisk, and
  Weitzman]{chunara2017denominator}
R.~Chunara, L.~E. Wisk, and E.~R. Weitzman.
\newblock Denominator issues for personally generated data in population health
  monitoring.
\newblock \emph{American journal of preventive medicine}, 52\penalty0
  (4):\penalty0 549--553, 2017.

\bibitem[Chunara et~al.(2020)Chunara, Plymoth, and Martin]{chunaradiversitysur}
R.~Chunara, A.~Plymoth, and L.~Martin.
\newblock Diversity in surveillance data: implications for infectious disease
  forecasting models.
\newblock \emph{Under review}, 2020.

\bibitem[Crenshaw(1989)]{crenshaw1989demarginalizing}
K.~Crenshaw.
\newblock Demarginalizing the intersection of race and sex: A black feminist
  critique of antidiscrimination doctrine, feminist theory and antiracist
  politics.
\newblock \emph{u. Chi. Legal f.}, page 139, 1989.

\bibitem[Daughton et~al.(2020)Daughton, Chunara, and
  Paul]{daughton2020comparison}
A.~R. Daughton, R.~Chunara, and M.~J. Paul.
\newblock Comparison of social media, syndromic surveillance, and microbiologic
  acute respiratory infection data: Observational study.
\newblock \emph{JMIR Public Health and Surveillance}, 6\penalty0 (2):\penalty0
  e14986, 2020.

\bibitem[Dugan et~al.(2015)Dugan, Mukhopadhyay, Carroll, and
  Downs]{dugan2015machine}
T.~M. Dugan, S.~Mukhopadhyay, A.~Carroll, and S.~Downs.
\newblock Machine learning techniques for prediction of early childhood
  obesity.
\newblock \emph{Applied clinical informatics}, 6\penalty0 (03):\penalty0
  506--520, 2015.

\bibitem[Eneanya et~al.(2019)Eneanya, Yang, and
  Reese]{eneanya2019reconsidering}
N.~D. Eneanya, W.~Yang, and P.~P. Reese.
\newblock Reconsidering the consequences of using race to estimate kidney
  function.
\newblock \emph{Jama}, 322\penalty0 (2):\penalty0 113--114, 2019.

\bibitem[Evans et~al.(2018)Evans, Williams, Onnela, and
  Subramanian]{evans2018multilevel}
C.~R. Evans, D.~R. Williams, J.-P. Onnela, and S.~Subramanian.
\newblock A multilevel approach to modeling health inequalities at the
  intersection of multiple social identities.
\newblock \emph{Social Science \& Medicine}, 203:\penalty0 64--73, 2018.

\bibitem[for Health Statistics~(US)(1994)]{national1994plan}
N.~C. for Health Statistics~(US).
\newblock \emph{Plan and operation of the third National Health and Nutrition
  Examination Survey, 1988-94}.
\newblock Number~32. National Ctr for Health Statistics, 1994.

\bibitem[Friedler et~al.(2016)Friedler, Scheidegger, and
  Venkatasubramanian]{friedler2016possibility}
S.~A. Friedler, C.~Scheidegger, and S.~Venkatasubramanian.
\newblock On the (im) possibility of fairness.
\newblock \emph{arXiv preprint arXiv:1609.07236}, 2016.

\bibitem[Galiatsatos et~al.(2019)Galiatsatos, Follin, Uradu, Alghanim, Daniel,
  Saria, Townsend, Sylvester, Chanmugam, and Chen]{galiatsatos2019association}
P.~Galiatsatos, A.~Follin, N.~Uradu, F.~Alghanim, Y.~Daniel, S.~Saria,
  J.~Townsend, C.~Sylvester, A.~Chanmugam, and E.~Chen.
\newblock The association between neighborhood socioeconomic disadvantage and
  readmissions for patients hospitalized with sepsis.
\newblock In \emph{C94. The Impact of Social Determinants in Pulmonary and
  Critical Care}, pages A5569--A5569. American Thoracic Society, 2019.

\bibitem[Glymour and Glymour(2014)]{glymour2014commentary}
C.~Glymour and M.~R. Glymour.
\newblock Commentary: race and sex are causes.
\newblock \emph{Epidemiology}, 25\penalty0 (4):\penalty0 488--490, 2014.

\bibitem[Grubbs(2020)]{grubbs2020precision}
V.~Grubbs.
\newblock Precision in gfr reporting: Let’s stop playing the race card, 2020.

\bibitem[Hamad et~al.(2019)Hamad, Nguyen, Bhattacharya, Glymour, and
  Rehkopf]{hamad2019educational}
R.~Hamad, T.~T. Nguyen, J.~Bhattacharya, M.~M. Glymour, and D.~H. Rehkopf.
\newblock Educational attainment and cardiovascular disease in the united
  states: A quasi-experimental instrumental variables analysis.
\newblock \emph{PLoS medicine}, 16\penalty0 (6):\penalty0 e1002834, 2019.

\bibitem[Heiman and Artiga(2015)]{heiman2015beyond}
H.~J. Heiman and S.~Artiga.
\newblock Beyond health care: the role of social determinants in promoting
  health and health equity.
\newblock \emph{Health}, 20\penalty0 (10):\penalty0 1--10, 2015.

\bibitem[Holstein et~al.(2019)Holstein, Wortman~Vaughan, Daum{\'e}~III, Dudik,
  and Wallach]{holstein2019improving}
K.~Holstein, J.~Wortman~Vaughan, H.~Daum{\'e}~III, M.~Dudik, and H.~Wallach.
\newblock Improving fairness in machine learning systems: What do industry
  practitioners need?
\newblock In \emph{Proceedings of the 2019 CHI Conference on Human Factors in
  Computing Systems}, pages 1--16, 2019.

\bibitem[Kasy and Abebe(2020)]{kasy2020fairness}
M.~Kasy and R.~Abebe.
\newblock Fairness, equality, and power in algorithmic decision making.
\newblock Technical report, Working paper, 2020.

\bibitem[Kreif et~al.(2015)Kreif, Grieve, D{\'\i}az, and
  Harrison]{kreif2015evaluation}
N.~Kreif, R.~Grieve, I.~D{\'\i}az, and D.~Harrison.
\newblock Evaluation of the effect of a continuous treatment: a machine
  learning approach with an application to treatment for traumatic brain
  injury.
\newblock \emph{Health economics}, 24\penalty0 (9):\penalty0 1213--1228, 2015.

\bibitem[Kusnoor et~al.(2018)Kusnoor, Koonce, Hurley, McClellan, Blasingame,
  Frakes, Huang, Epelbaum, and Giuse]{kusnoor2018collection}
S.~V. Kusnoor, T.~Y. Koonce, S.~T. Hurley, K.~M. McClellan, M.~N. Blasingame,
  E.~T. Frakes, L.-C. Huang, M.~I. Epelbaum, and N.~B. Giuse.
\newblock Collection of social determinants of health in the community clinic
  setting: A cross-sectional study.
\newblock \emph{BMC Public Health}, 18\penalty0 (1):\penalty0 550, 2018.

\bibitem[Lazer et~al.(2014)Lazer, Kennedy, King, and
  Vespignani]{lazer2014parable}
D.~Lazer, R.~Kennedy, G.~King, and A.~Vespignani.
\newblock The parable of google flu: traps in big data analysis.
\newblock \emph{Science}, 343\penalty0 (6176):\penalty0 1203--1205, 2014.

\bibitem[L{\^e}-Scherban et~al.(2018)L{\^e}-Scherban, Wang, Boyle-Steed, and
  Pachter]{le2018intergenerational}
F.~L{\^e}-Scherban, X.~Wang, K.~H. Boyle-Steed, and L.~M. Pachter.
\newblock Intergenerational associations of parent adverse childhood
  experiences and child health outcomes.
\newblock \emph{Pediatrics}, 141\penalty0 (6):\penalty0 e20174274, 2018.

\bibitem[Lehman(2019)]{lehman2019addressing}
C.~Lehman.
\newblock Addressing social determinants of health.
\newblock \emph{Physical Therapy in Motion}, 2019.

\bibitem[Levey et~al.(2020)Levey, Titan, Powe, Coresh, and
  Inker]{levey2020kidney}
A.~S. Levey, S.~M. Titan, N.~R. Powe, J.~Coresh, and L.~A. Inker.
\newblock Kidney disease, race, and gfr estimation.
\newblock \emph{Clinical Journal of the American Society of Nephrology}, 2020.

\bibitem[Lodi et~al.(2019)Lodi, Phillips, Lundgren, Logan, Sharma, Cole,
  Babiker, Law, Chu, Byrne, et~al.]{lodi2019effect}
S.~Lodi, A.~Phillips, J.~Lundgren, R.~Logan, S.~Sharma, S.~R. Cole, A.~Babiker,
  M.~Law, H.~Chu, D.~Byrne, et~al.
\newblock Effect estimates in randomized trials and observational studies:
  comparing apples with apples.
\newblock \emph{American journal of epidemiology}, 188\penalty0 (8):\penalty0
  1569--1577, 2019.

\bibitem[Mahamoud et~al.(2013)Mahamoud, Roche, and
  Homer]{mahamoud2013modelling}
A.~Mahamoud, B.~Roche, and J.~Homer.
\newblock Modelling the social determinants of health and simulating short-term
  and long-term intervention impacts for the city of toronto, canada.
\newblock \emph{Social science \& medicine}, 93:\penalty0 247--255, 2013.

\bibitem[McCradden et~al.(2020)McCradden, Joshi, Mazwi, and
  Anderson]{mccradden2020ethical}
M.~D. McCradden, S.~Joshi, M.~Mazwi, and J.~A. Anderson.
\newblock Ethical limitations of algorithmic fairness solutions in health care
  machine learning.
\newblock \emph{The Lancet Digital Health}, 2\penalty0 (5):\penalty0
  e221--e223, 2020.

\bibitem[Mhasawade et~al.(2020{\natexlab{a}})Mhasawade, Elghafari, Duncan, and
  Chunara]{mhasawade2020role}
V.~Mhasawade, A.~Elghafari, D.~T. Duncan, and R.~Chunara.
\newblock Role of the built and online social environments on expression of
  dining on instagram.
\newblock \emph{International journal of environmental research and public
  health}, 17\penalty0 (3):\penalty0 735, 2020{\natexlab{a}}.

\bibitem[Mhasawade et~al.(2020{\natexlab{b}})Mhasawade, Rehman, and
  Chunara]{mhasawade2020population}
V.~Mhasawade, N.~A. Rehman, and R.~Chunara.
\newblock Population-aware hierarchical bayesian domain adaptation via
  multi-component invariant learning.
\newblock In \emph{Proceedings of the ACM Conference on Health, Inference, and
  Learning}, pages 182--192, 2020{\natexlab{b}}.

\bibitem[Mitchell and Jolley(2004)]{mitchell2004research}
M.~Mitchell and J.~Jolley.
\newblock Research design explained 5 th ed.
\newblock \emph{Victoria: Wadsworth Publisher. Moebert, J. \& Tydecks,
  P.(2007). Power and Ownership Structures among German Companies. A Network
  Analysis of Financial Linkages}, 2004.

\bibitem[Mitchell et~al.(2018)Mitchell, Potash, Barocas, D'Amour, and
  Lum]{mitchell2018prediction}
S.~Mitchell, E.~Potash, S.~Barocas, A.~D'Amour, and K.~Lum.
\newblock Prediction-based decisions and fairness: A catalogue of choices,
  assumptions, and definitions.
\newblock \emph{arXiv preprint arXiv:1811.07867}, 2018.

\bibitem[Mooney and Pejaver(2018)]{mooney2018big}
S.~J. Mooney and V.~Pejaver.
\newblock Big data in public health: terminology, machine learning, and
  privacy.
\newblock \emph{Annual review of public health}, 39:\penalty0 95--112, 2018.

\bibitem[Morland and Evenson(2009)]{morland2009obesity}
K.~B. Morland and K.~R. Evenson.
\newblock Obesity prevalence and the local food environment.
\newblock \emph{Health \& place}, 15\penalty0 (2):\penalty0 491--495, 2009.

\bibitem[{National Academies}(2015)]{national2015growing}
{National Academies}.
\newblock \emph{The growing gap in life expectancy by income: Implications for
  federal programs and policy responses}.
\newblock National Academies Press, 2015.

\bibitem[Nishtala et~al.(2020)Nishtala, Kamarthi, Thakkar, Narayanan, Grama,
  Padmanabhan, Madhiwalla, Chaudhary, Ravindra, and Tambe]{nishtala2020missed}
S.~Nishtala, H.~Kamarthi, D.~Thakkar, D.~Narayanan, A.~Grama, R.~Padmanabhan,
  N.~Madhiwalla, S.~Chaudhary, B.~Ravindra, and M.~Tambe.
\newblock Missed calls, automated calls and health support: Using ai to improve
  maternal health outcomes by increasing program engagement.
\newblock \emph{arXiv preprint arXiv:2006.07590}, 2020.

\bibitem[Nolen(2020)]{raceAKI}
L.~Nolen.
\newblock \emph{Elimination of race coefficient from eGFR calculation}, 2020.
\newblock URL \url{https://twitter.com/LashNolen/status/1276181898394558467}.

\bibitem[Obermeyer and Mullainathan(2019)]{obermeyer2019dissecting}
Z.~Obermeyer and S.~Mullainathan.
\newblock Dissecting racial bias in an algorithm that guides health decisions
  for 70 million people.
\newblock In \emph{Proceedings of the Conference on Fairness, Accountability,
  and Transparency}, pages 89--89, 2019.

\bibitem[O'Connor and Matthews(2011)]{o2011informational}
J.~O'Connor and G.~Matthews.
\newblock Informational privacy, public health, and state laws.
\newblock \emph{American journal of public health}, 101\penalty0 (10):\penalty0
  1845--1850, 2011.

\bibitem[Penman-Aguilar et~al.(2016)Penman-Aguilar, Talih, Huang, Moonesinghe,
  Bouye, and Beckles]{penman2016measurement}
A.~Penman-Aguilar, M.~Talih, D.~Huang, R.~Moonesinghe, K.~Bouye, and
  G.~Beckles.
\newblock Measurement of health disparities, health inequities, and social
  determinants of health to support the advancement of health equity.
\newblock \emph{Journal of public health management and practice: JPHMP},
  22\penalty0 (Suppl 1):\penalty0 S33, 2016.

\bibitem[Quisel et~al.(2016)Quisel, Kale, and Foschini]{quisel2016intra}
T.~Quisel, D.~C. Kale, and L.~Foschini.
\newblock Intra-day activity better predicts chronic conditions.
\newblock \emph{arXiv preprint arXiv:1612.01200}, 2016.

\bibitem[Rahmattalabi et~al.(2018)Rahmattalabi, Adhikari, Vayanos, Tambe, Rice,
  and Baker]{rahmattalabi2018influence}
A.~Rahmattalabi, A.~B. Adhikari, P.~Vayanos, M.~Tambe, E.~Rice, and R.~Baker.
\newblock Influence maximization for social network based substance abuse
  prevention.
\newblock In \emph{Thirty-Second AAAI Conference on Artificial Intelligence},
  2018.

\bibitem[Rajkomar et~al.(2018{\natexlab{a}})Rajkomar, Hardt, Howell, Corrado,
  and Chin]{rajkomar2018ensuring}
A.~Rajkomar, M.~Hardt, M.~D. Howell, G.~Corrado, and M.~H. Chin.
\newblock Ensuring fairness in machine learning to advance health equity.
\newblock \emph{Annals of internal medicine}, 169\penalty0 (12):\penalty0
  866--872, 2018{\natexlab{a}}.

\bibitem[Rajkomar et~al.(2018{\natexlab{b}})Rajkomar, Oren, Chen, Dai, Hajaj,
  Hardt, Liu, Liu, Marcus, Sun, et~al.]{rajkomar2018scalable}
A.~Rajkomar, E.~Oren, K.~Chen, A.~M. Dai, N.~Hajaj, M.~Hardt, P.~J. Liu,
  X.~Liu, J.~Marcus, M.~Sun, et~al.
\newblock Scalable and accurate deep learning with electronic health records.
\newblock \emph{NPJ Digital Medicine}, 1\penalty0 (1):\penalty0 18,
  2018{\natexlab{b}}.

\bibitem[Rose(1985)]{rose1985sick}
G.~Rose.
\newblock Sick individuals and sick populations, int. journ.
\newblock \emph{Of Epidemiology}, 14\penalty0 (1), 1985.

\bibitem[Snyder et~al.(2018)Snyder, Salkowski, Wey, Pyke, Israni, and
  Kasiske]{snyder2018organ}
J.~J. Snyder, N.~Salkowski, A.~Wey, J.~Pyke, A.~K. Israni, and B.~L. Kasiske.
\newblock Organ distribution without geographic boundaries: a possible
  framework for organ allocation.
\newblock \emph{American Journal of Transplantation}, 18\penalty0
  (11):\penalty0 2635--2640, 2018.

\bibitem[Stein et~al.(1993)Stein, Lederman, and Shea]{stein1993behavioral}
A.~D. Stein, R.~I. Lederman, and S.~Shea.
\newblock The behavioral risk factor surveillance system questionnaire: its
  reliability in a statewide sample.
\newblock \emph{American Journal of Public Health}, 83\penalty0 (12):\penalty0
  1768--1772, 1993.

\bibitem[Stern et~al.(2016)Stern, Poti, Ng, Robinson, Gordon-Larsen, and
  Popkin]{stern2016people}
D.~Stern, J.~M. Poti, S.~W. Ng, W.~R. Robinson, P.~Gordon-Larsen, and B.~M.
  Popkin.
\newblock Where people shop is not associated with the nutrient quality of
  packaged foods for any racial-ethnic group in the united states.
\newblock \emph{The American journal of clinical nutrition}, 103\penalty0
  (4):\penalty0 1125--1134, 2016.

\bibitem[VanderWeele and Robinson(2014)]{vanderweele2014causal}
T.~J. VanderWeele and W.~R. Robinson.
\newblock On causal interpretation of race in regressions adjusting for
  confounding and mediating variables.
\newblock \emph{Epidemiology (Cambridge, Mass.)}, 25\penalty0 (4):\penalty0
  473, 2014.

\bibitem[Vyas et~al.(2020)Vyas, Eisenstein, and Jones]{vyas2020hidden}
D.~A. Vyas, L.~G. Eisenstein, and D.~S. Jones.
\newblock Hidden in plain sight—reconsidering the use of race correction in
  clinical algorithms, 2020.

\bibitem[WHO(2008)]{who2008closing}
WHO.
\newblock \emph{Closing the gap in a generation: Health equity through action
  on the social determinants of health: Commission on Social Determinants of
  Health final report}.
\newblock World Health Organization, 2008.

\bibitem[{WHO}(2009)]{world20092008}
{WHO}.
\newblock 2008-2013 action plan for the global strategy for the prevention and
  control of noncommunicable diseases: prevent and control cardiovascular
  diseases, cancers, chronic respiratory diseases and diabetes.
\newblock 2009.

\bibitem[Wu et~al.(2019)Wu, Zhang, Wu, and Tong]{wu2019pc}
Y.~Wu, L.~Zhang, X.~Wu, and H.~Tong.
\newblock Pc-fairness: A unified framework for measuring causality-based
  fairness.
\newblock In \emph{Advances in Neural Information Processing Systems}, pages
  3404--3414, 2019.

\bibitem[Yang et~al.(2020)Yang, Loftus, and Stoyanovich]{yang2020causal}
K.~Yang, J.~R. Loftus, and J.~Stoyanovich.
\newblock Causal intersectionality for fair ranking.
\newblock \emph{arXiv preprint arXiv:2006.08688}, 2020.

\bibitem[Zhan et~al.(2018)Zhan, Mohan, Tarolli, Schneider, Adams, Sharma,
  Elson, Spear, Glidden, Little, et~al.]{zhan2018using}
A.~Zhan, S.~Mohan, C.~Tarolli, R.~B. Schneider, J.~L. Adams, S.~Sharma, M.~J.
  Elson, K.~L. Spear, A.~M. Glidden, M.~A. Little, et~al.
\newblock Using smartphones and machine learning to quantify parkinson disease
  severity: the mobile parkinson disease score.
\newblock \emph{JAMA neurology}, 75\penalty0 (7):\penalty0 876--880, 2018.

\bibitem[Zhao et~al.(2020)Zhao, Mirin, Wood, Rajesh, Cook, and
  Chunara]{yuan2020machine}
Y.~Zhao, N.~Mirin, E.~Wood, V.~Rajesh, S.~Cook, and R.~Chunara.
\newblock Machine learning for integrating social determinants in
  cardiovascular disease prediction models: A systematic review.
\newblock \emph{In submission}, 2020.

\bibitem[Zoler(2020)]{raceAKIdrop}
M.~L. Zoler.
\newblock \emph{Dropping Race-Based eGFR Adjustment Gains Traction in US},
  2020.
\newblock URL \url{https://www.medscape.com/viewarticle/933418}.

\end{thebibliography}
\clearpage

\end{document}